\documentclass[sigconf]{acmart}
\AtBeginDocument{%
  \providecommand\BibTeX{{%
    \normalfont B\kern-0.5em{\scshape i\kern-0
    .25em b}\kern-0.8em\TeX}}}

\copyrightyear{2022} 
\acmYear{2022} 
\setcopyright{rightsretained} 
\acmConference[MODELS '22 Companion]{ACM/IEEE 25th International Conference on Model Driven Engineering Languages and Systems}{October 23--28, 2022}{Montreal, QC, Canada}
\acmBooktitle{ACM/IEEE 25th International Conference on Model Driven Engineering Languages and Systems (MODELS '22 Companion), October 23--28, 2022, Montreal, QC, Canada}
\acmDOI{10.1145/3550356.3561595}
\acmISBN{978-1-4503-9467-3/22/10}

\usepackage{todonotes}

\usepackage{paralist}
\usepackage{listings}
\usepackage{hyperref}
\usepackage{dirtree}

\newtheorem{Definition}{Definition}

\newcommand{\quotes}[1]{``#1''}

\hypersetup{breaklinks=true}
\begin{document}

\pagestyle{plain}

\title{Design Guidelines for Improving User Experience in Industrial Domain-Specific Modelling Languages}


\author{Rohit Gupta}
\affiliation{%
  \institution{Siemens AG}
  \city{Munich}
  \country{Germany}}
\email{rg.gupta@siemens.com}

\author{Nico Jansen}
\affiliation{%
  \institution{Software Engineering\\RWTH Aachen University}
  \city{Aachen}
  \country{Germany}}
\email{jansen@se-rwth.de}

\author{Nikolaus Regnat}
\affiliation{%
  \institution{Siemens AG}
  \city{Munich}
  \country{Germany}}
\email{nikolaus.regnat@siemens.com}

\author{Bernhard Rumpe}
\affiliation{%
  \institution{Software Engineering\\RWTH Aachen University}
  \city{Aachen}
  \country{Germany}}
\email{rumpe@se-rwth.de}

\begin{abstract}
Domain-specific modelling languages (DSMLs) help practitioners solve modelling challenges specific to various domains.
As domains grow more complex and heterogeneous in nature, industrial practitioners often face challenges in the usability of graphical DSMLs.
There is still a lack of guidelines that industrial language engineers should consider for improving the user experience (UX) of these practitioners.
The overall topic of UX is vast and subjective, and general guidelines and definitions of UX are often overly generic or tied to specific technological spaces.
To solve this challenge, we leverage existing design principles and standards of human-centred design and UX in general and propose definitions and guidelines for UX and user experience design (UXD) aspects in graphical DSMLs. 
In this paper, we categorize the key UXD aspects, primarily based on our experience in developing industrial DSMLs, that language engineers should consider during graphical DSML development.
Ultimately, these UXD guidelines help to improve the general usability of industrial DSMLs and support language engineers in developing better DSMLs that are independent of graphical modelling tools and more widely accepted by their users.
\end{abstract}

\ccsdesc[500]{Software and its engineering~Domain specific languages}
\ccsdesc[500]{Software and its engineering~Model-driven software engineering}

\keywords{Domain-Specific Languages, Model-Based Systems Engineering, Industrial Domain-Specific Modelling Languages, Industrial Language Engineering, User Experience}

\maketitle

\vspace{-5mm}
\section{Introduction}
\label{sec:intro}


With the advancement of various systems engineering domains, we are observing a notable shift in the way modelling is being introduced in projects early on.
The engineering process in modelling involves models instead of documents for complex heterogeneous systems.
Consequently, model-based development and model-based systems engineering (MBSE) techniques are constantly evolving~\cite{FR07}.
General-Purpose Languages (GPLs) often provide difficulties in system modelling~\cite{DBLP:journals/emisa/ProperB19}, most notably by ignoring domain experts from contributing to solutions.
Instead, Domain-Specific Languages (DSLs) aim to reduce the gap in a particular domain by supporting domain-specific abstractions~\cite{CBCR15,GKR+21}.
The technological spaces for DSLs are heterogeneous, either textual, graphical, or projectional~\cite{DCB+15, Bet16, tolvanen2006metaedit+}.
Regardless of their technological spaces, various domain-specific modelling languages (DSMLs) have been developed to support modelling in their respective domains.
Naturally, as the complexity of these DSMLs increases, so does the complexity in its engineering, as does its consequence on users.
This often leads to users, who may or may not be modelling experts, struggling to use DSMLs effectively. 


Modelling involves key decision making in systems engineering.
To assist end users in achieving their modelling goals, it is important for the DSML to convey all relevant aspects of their domain.
The entire lifecycle involved in the modelling process should not only involve an elaborate modelling language, but also a solid methodological foundation and an appropriate tooling mechanism~\cite{GKR+21} (\autoref{sec:modInMD}).
Only with the combination of these aspects would modelling be effective with novice and advanced users and with small and medium enterprises~\cite{DBLP:conf/modellierung/Regnat18}.
Therefore, as domains become more heterogeneous, the complexity of the syntax and the semantics of the language increases~\cite{CBCR15, CGR09}, and the need for a good user experience (UX) also grows.
Aspects of UX are often overlooked by language engineers because of the false notion that users are always modelling experts in their domains and can understand every construct of a language easily.
Especially in industrial DSMLs, novice users who are introduced to graphical modelling tools, often in the latter stages of the project lifecycle, need guidance because of their lack of know-how in modelling.
Challenges of UX in using graphical DSMLs include, but are not limited to, improper visual representation of domain aspects, difficult to find and use language elements, burdening users with unnecessary tool functionalities~\cite{whittle2013industrial}, unavailability of predefined templates of models, and unclear documentation for the modelling language constructs.

Various definitions of UX and usability heuristics have been proposed in the literature to solve these challenges~\cite{10.1109/TSE.2009.67,10.1145/1512714.1512717}.
However, these are overly generic and often tied to specific technological spaces.
Further, UX is a subjective topic, the preference of users vastly differs across domains and is influenced by recent trends from the ubiquitous software domain (\autoref{sec:bg}).
UX and business logic is often intertwined in projects, and while considerations are made by involving key stakeholders early on, they are often lost in translation as the project complexity grows.
Naturally, no cookie cutter solutions exist, but there are industry standards, design guidelines~\cite{KKP+09,czech2018best,voelterbest}, and user experience design (UXD) aspects that language engineers should consider when developing graphical DSMLs.
Improving UX for users is key to achieving modelling goals.
To this end, we aim to provide guidance based on our experiences in developing industrial DSMLs combining aspects of UX for a more holistic modelling experience for users of various domains. 


Given our years of experience in building industrial DSMLs for a variety of users in various domains using graphical modelling tools such as Enterprise Architect~\cite{EnterpriseArchitect}, Rational Rhapsody~\cite{IBMRhapsody}, and MagicDraw~\cite{MagicDraw}, we put the choice on a firmer basis in this paper:
\begin{itemize}
	\itemsep0em
	\item We define the terms \textit{user experience (UX)} and \textit{user experience design (UXD)} for graphical DSMLs in general modelling tools on the example of MagicDraw (\autoref{sec:uxdInMD}).
	\item We discuss human-centred design approaches and define the important UXD aspects: visual design, information architecture, interaction design, and usability heuristics that language engineers should consider during graphical DSML development. 
	\item We illustrate the UXD aspects using an example of a feature model DSML with product lines and products (\autoref{sec:impl}).
	\item Finally, we clarify the scope of UX in our modelling approaches,  discuss the benefits of our approach (\autoref{sec:discussion}), related work (\autoref{sec:relWork}), and conclude the paper (\autoref{sec:conclusions}).
\end{itemize}

\section{Background}
\label{sec:bg}

The engineering of DSLs and DSMLs comes with the usual challenges faced in software engineering, such as maintenance or evolution, as they are software themselves~\cite{FGD+10}.
Generally speaking, a software language consists of~\cite{CGR09,CBCR15}:
\begin{inparaenum}[(1)]
	\item an abstract syntax that defines the structure of its models, e.g., in 
	form of context-free grammars~\cite{HKR21};
	\item a concrete syntax that defines how the models are presented, e.g., 
	graphical~\cite{DCB+15}, textual~\cite{Bet16}, or projectional~\cite{Cam14},
	\item semantics, in the sense of meaning~\cite{HR04}; and
	\item context conditions, to check the well-formedness of a language.
\end{inparaenum}
Methodologies for developing graphical DSMLs in the industry are often tied to specific departments, where language engineers must be trained to represent the domain in consideration effectively.
The development of such graphical DSMLs must also include all functional and non-functional aspects of the language.
Only with the combination of a modelling language, a methodology, and a modelling tool can this challenge be fully addressed.
Central research units such as Siemens Technology, therefore, provide guidelines to their language engineers in designing DSMLs that improve the overall UX for both novice and advanced modellers across various domains.

Several definitions of UX~\cite{10.1145/1512714.1512717} and best practices for DSLs and model-driven development~\cite{voelterbest} exist in general.
Consequently, a number of notions, such as feelings, experiences, and insights, are subsumed under the definitions of UX.
This could be attributed to the fact that UX is a vastly subjective topic covering a multitude of domains, one that cannot be possibly represented by a single definition.
ISO 9241-210~\cite{ISO9241210} defines UX as \quotes{person's perceptions and responses resulting from the use and/or anticipated use of a product, system or service}.
While such a definition can be reused in the context of graphical modelling, it is overly generic, needs more clarity, and does not discuss domain-specific aspects.
There is also a lack of consensus as to which UX and usability definitions (e.g., ISO 9241-11~\cite{bevan2015iso}, ISO 13407~\cite{10.1145/944519.944525}, Nielsen's~\cite{nielsen}) are best suited for graphical DSMLs.
Aspects of Human-Computer Interaction (HCI) discusses how people interface with computers (e.g., ISO 9241-161~\cite{ISO9241161}) and how it can be leveraged to obtain practical results related to user interfaces~\cite{poltronieri2017usability}.
The proposed definitions of UX, usability, and HCI are often overlooked by language engineers~\cite{abrahao2017user}, who are often neither UX experts nor domain experts in various domains.
Thus, developing graphical DSMLs requires focusing on specific aspects of UX, and that a one-size-fits-all approach is probably not suited. 

A good UX is key to modellers effectively reaching their modelling goals.
Benefits of a good UX in modelling ameliorate the challenges between the users and the constructs of the language and reduce unnecessary burdens of the modelling tool functionalities.
In our experience of developing a diverse set of industrial graphical DSMLs, we have not come across a standard set of UX and usability guidelines that language engineers should consider during graphical DSMLs development.
We believe there is a need to start the discussion towards defining specific UX guidelines that benefit novice and advanced users and is ultimately independent of a specific implementation or a graphical modelling tool.
\vspace{-4pt}

\section{Modelling in MagicDraw}
\label{sec:modInMD}

The methodologies for the development of graphical DSMLs are generally tied to specific departments in a large organization such as Siemens AG.
This introduces challenges in the way the combination of a modelling language is used with a methodology and using an appropriate graphical modelling tool.
We have focussed on MagicDraw as a choice of modelling tool in developing graphical DSMLs, as it is based on the Unified Modelling Language (UML), comes with comprehensive extensions such as the Systems Modelling Language (SysML) plugin, and provides a wide range of customization possibilities that are used to capture most, if not all, issues of domain-specific language engineering.
A systematic engineering process of developing industrial DSLs~\cite{fowler2010domain} using modular reusable DSL Building Blocks in MagicDraw is described in~\cite{GKR+21}.
Each building block conceptually consists of language components, a method and a user experience design (UXD) part.
The re-usable language components define the language~\cite{Rum16}, wholly or in part.
The method part describes a suitable methodology for the language to help users achieve their modelling goals.
This can be in the form of training material, including methodical steps, that ultimately provide a comprehensive guidance to DSML users.
Finally, the UXD part describes the design decisions language engineers must consider to improve the overall usability and UX of a DSL or DSML.
The heterogeneous building blocks are finally composed together to create the DSL.
The combination of the modelling language, method, and design decisions makes modelling effective and simpler for all kinds of users, novice and advanced alike.
To this end, we consider UX aspects in a DSL as important as the modelling language itself.

\begin{figure*}[h!]
	\centering
	\includegraphics[width=\textwidth]{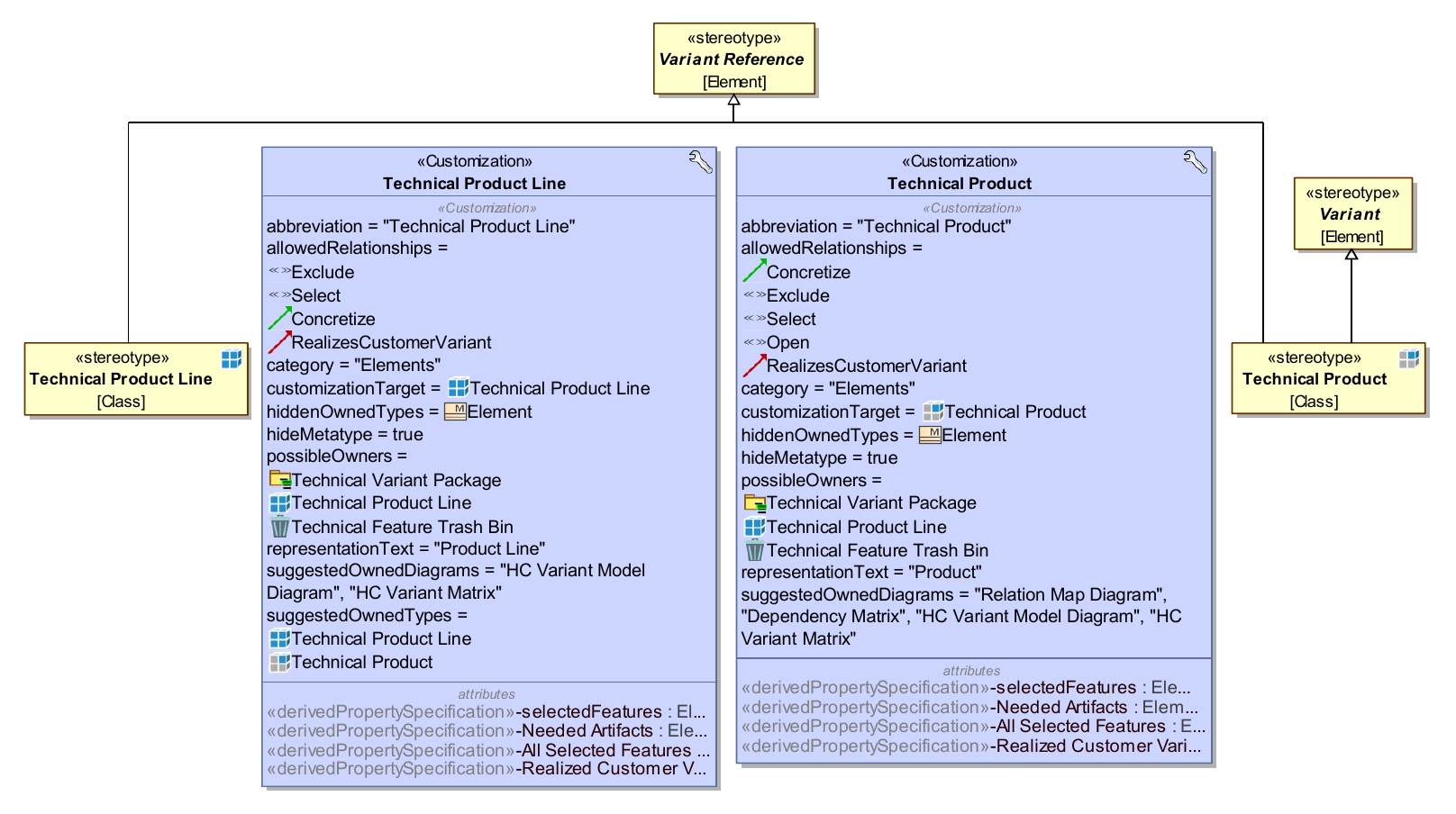}
	\caption{An example of the configuration of the product line and product language elements for a feature model DSL using the customization capabilities of MagicDraw.}
	\label{fig:variantConfig}
\end{figure*}

The development process is iterative and involves all key stakeholders in the project.
A feedback loop is established between modellers, domain experts, and the language engineers to ensure aspects of the domain are continuously integrated in the DSML.
Further, UX experts are involved to ensure aspects of user experience and usability are effectively captured in graphical DSMLs.
The customization capabilities of MagicDraw allows creation of a language profile consisting of language component artefacts.
An example of an artefact is a language element, referred to as a \textit{stereotype} in MagicDraw.
In addition, \textit{customization} elements allow the definition of rules for the MagicDraw DSL customization engine. 
\autoref{fig:variantConfig} shows an example of the configuration of stereotypes for a feature model~\cite{BEK+19} along with its relationships and customizations in MagicDraw.
These customizations define what model elements or diagrams can be created and where, how an element shortcut menu should look like, what properties of the elements are presented to the user, and so on.   
MagicDraw also provides a specific Java based plugin mechanism that supports the integration of automation and creation of custom functionalities.
Additionally, model templates allow a predefined model structure to be automatically instantiated during modelling and perspectives help configure the visible number of functionalities of MagicDraw for different kinds of users. 


Over the years, we have used this methodology to build industrial DSMLs in MagicDraw for a variety of consumers in various domains.
In the case of Siemens Healthineers, we developed DSMLs that focussed on the modelling of medical devices, imaging for radiation therapy, and X-ray equipment.
For Siemens IT, we developed a DSML that models the concerned IT workflow and inventory management processing.
For Siemens Energy AG, we developed DSMLs that allow for the modelling of complex industrial appliances such as turbines along with their workflows.
In the case of Siemens Digital Industry, we developed DSMLs to model the hardware and software aspects of the SINAMICS\footnote{\url{https://new.siemens.com/global/en/products/drives/sinamics.html}} frequency converter that allows for variable frequency drives and control systems, whereas the SIMOTICS\footnote{\url{https://new.siemens.com/global/en/products/drives/electric-motors.html}} electric motors allow the modelling of synchronous as well as asynchronous industrial motors.
In a recently ongoing public funded SpesML project~\cite{mci/Regnat2022}, we built modular reusable DSL Building Blocks to separate the concerns of the models created for a system under development as different heterogeneous viewpoints.
In all of these projects, feedback from users regarding the usability and UX of the DSMLs, to invoke positive feelings during the interaction with the DSMLs, have been of paramount importance.

\section{User Experience in MagicDraw}
\label{sec:uxdInMD}

\begin{Definition}[User Experience in MagicDraw]
	We define user experience (UX) for graphical DSMLs in MagicDraw as an instantaneous intuitive feeling (positive or negative) of a user (modeller) while interacting with the defined constructs of the graphical modelling language and the graphical modelling tool, MagicDraw.
\end{Definition}


In other words, UX is primarily an 
intuitive 
feeling for users during modelling with the aim that a good UX satisfies their modelling expectations in easy, simple terms while also minimising the number of interactions required between them and the modelling tool.
These interactions are the abilities of systems and users to influence each other in order to reach a goal.
We consider any positive feeling during the interaction with the modelling language and the modelling tool as a good UX.
A bad UX generally tends to invoke negative feelings that not only leaves DSML users dissatisfied, but also introduces incomprehensibility between the different stakeholders in their modelling.
\begin{Definition}[User Experience Design in MagicDraw]
	We define user experience design (UXD) as any design decision taken by a language engineer during the development of a graphical DSML in MagicDraw, that ultimately fosters a good UX for a user.
\end{Definition}

The design decisions are realized and implemented by language engineers in consultation with practitioners during the graphical DSML development process.
Any design decision should follow the principles of human-centred design as defined in ISO 9241-210.
Language engineers must take into account the people who use the graphical modelling language as well as other stakeholders who are involved in the project.
The following categorization of UXD aspects in MagicDraw originate from experiences found in the literature~\cite{garrett2010elements}.
These UXD aspects are non-exhaustive but provide a general foundation for graphical DSMLs, independent of their modelling tools.
Each of these design decisions is elaborated with a rationale that provides a reasoning as to why the design decision is required and consequently what its benefits are.
These rationales are also based on feedback from users and domain experts in various projects (\autoref{sec:modInMD}).
We also assign an identifier (in brackets) to distinguish each design decisions based on their categories.

\subsection{Categorization of UXD}
\subsubsection{Visual Design}
Visual designs represent the aesthetics or the look and feel of models and model elements and how they are presented to users~\cite{10.1109/TSE.2009.67}.
Model elements can be configured in the form of various icons, colours, appearance, dialogs along with their properties such as shape, size, and opacity.
In other words, the graphical concrete syntax of the DSML is enhanced using the following visual designs to effectively represent the various heterogeneous domains involved in modelling:
\begin{itemize}
	\itemsep0em 
	\item \textbf{Icon (V-1)}: an extra graphical element that is displayed upon selection for a specific model element.
	
	\textit{Rationale}: Icons help to distinguish between different model elements and convey real-world representations of a specific entity with some meaning for a user.
	
	\item \textbf{Colour (V-2)}: enhances the appearance of a model element through a specific colour.
	
	\textit{Rationale}: Colouring schemes such as red for errors or warnings provoke an increased user attention~\cite{brave2007emotion} and subsequently aids in differentiating model elements.
	
	\item \textbf{Modal Dialog (V-3)}: a graphical control element showing information to users on making relevant modelling decisions.

	\textit{Rationale}: Modal dialogs direct important information towards users and therefore require a user's immediate attention. Although it is useful in preventing or correcting vital errors, a drawback of a modal dialog is that it often interrupts a user's workflow.
	
	\item \textbf{Custom View (V-4)}: a visual representation of the textual information of model elements in the form of matrices, tables, UML or free-form diagrams.
	
	\textit{Rationale}: Custom views represent textual information visually using specific diagrams that contains information in a particular format. As each diagram serves a specific purpose, language engineers must consider the suitability for such diagrams and implement them in the language definition.
	
	\item \textbf{Dynamic View Plugin (V-5)}: a GPL (e.g., Java) based plugin for enabling dynamic filtering and/or displaying specific model information (e.g., legends, annotations) on existing UML diagrams or custom views.
	
	\textit{Rationale}: Dynamic view plugins help users focus on specific model information, for example, supporting the enabling or disabling of a power supply unit in a complex power system. Other model information such as legends or annotations from other users can also be toggled using a dynamic view plugin, allowing quick display or removal of information on different views based on certain filtering conditions.
	
\end{itemize}

\subsubsection{Information Architecture}
Information architecture is the practice of structuring and organizing the constructs of the graphical DSML in a way that they are easy to find and use~\cite{de2016user}.
As heterogeneous systems grow in complexity, so does the complexity in modelling.
With a growing number of domain concepts and functionalities that a graphical modelling tool provides, there is a need to alleviate the concerns of users, novice or advanced, for presenting the DSML and the functionalities of MagicDraw in a structured and organized manner.
We consider the following information architecture designs important in graphical modelling:
\begin{itemize}
	\itemsep0em 
	\item \textbf{Layout (IA-1)}: determination of the position of model elements on custom diagrams based on the context of use.
	
	\textit{Rationale}: The placement of model elements on specific areas of custom diagrams helps provide structure in complex modelling scenarios.
	An example is the positioning of incoming ports, generally to the left of a model element, or outgoing ports, which are positioned to the right of a model element.
	
	\item \textbf{Model Browser (IA-2)}: a visual representation of the hierarchy of model elements. It is a hierarchical navigation tool for managing the model data.
	
	\textit{Rationale}: Users often struggle to quickly navigate, find, or arrange model elements, especially in complex DSMLs. A model browser is designed to provide a sound hierarchical structure and organization of such model elements.
	
	\item \textbf{Perspective (IA-3)}: displaying a fixed set of modelling language constructs or tool functionalities based on the kind of user, novice or advanced.
	
	\textit{Rationale}: Often, novice users do not need additional tooling functionalities or language constructs to start their modelling. As the complexity of models increases, users tend to use more advanced concepts that require the enabling of advanced functionalities.
	
	\item \textbf{Creation View (IA-4)}: an additional pane or window that shows the logical grouping of different language elements, standard UML diagrams and custom views during model element creation on the model browser.
	
	\textit{Rationale}: Language elements that belong to a certain logical grouping should not belong to other logical groups to avoid disorder and inconsistencies in structuring and organizing the constructs of the modelling language. As an example, model elements and custom views of functional and non-functional requirements should exclusively belong to a logical group that defines a requirements specification.
	
\end{itemize}

\subsubsection{Interaction Design}
Interaction designs in MagicDraw help users interact effectively with the constructs of the graphical DSML while focussing on the cognitive dimensions~\cite{nielsen1994enhancing,blackwell2001cognitive}.
Such designs subsumes auto-completion of model element names, automatic instantiation of model elements, and transformation of models into other formalisms~\cite{kuhne2009systematic}.
Interaction designs focus mostly on the behaviour of the modelling language constructs.
We consider the following interaction designs important in graphical modelling:
\begin{itemize}
	\itemsep0em
	\item \textbf{Project Template (ID-1)}: is a customized project pattern that serves as a starting point for creating a project in a predefined format. 
	
	\textit{Rationale}: When users start modelling for the first time, they often lack the know-how on creating a robust, well-organized model consisting of possible model elements. Project templates help such users with the automatic instantiation of a model in a predefined format. An example is the automatic creation of a basic traffic light model as states in a state machine~\cite{GKR+21}, with only the triggers or actions that need to be additionally defined by the user.
	
	\item \textbf{Default Naming Scheme (ID-2)}: a naming scheme automatically assigning default names or numbers to model elements. 
	
	\textit{Rationale}: Users often forget to assign relevant names, labels, or numbering to model element to help them distinguish from other model elements and to avoid naming conflicts. In complex systems, this problem cascades drastically and adds unnecessary debug times. An example of a default naming scheme is to create a functional requirement as \quotes{Functional Requirement x}, x being an automatic incremental number.
	
	\item \textbf{Model Transformation (ID-3)}: a transformation of a model into another formalism.
	
	\textit{Rationale}: Model transformations are an effective way to dynamically transform models into other formalisms during design time. An example of a model transformation is to transform a model from an optional feature to a mandatory feature in a feature model during modelling. The user should be able to simply achieve such a transformation.
	
	\item \textbf{Custom User Interface (ID-4)}: a custom user interface (UI) programmed using frontend frameworks (e.g., Java Swing) to access and/or edit specific model and model elements based on the DSML requirements.
	
	\textit{Rationale}: Some modelling tools may not provide out of the box functionalities to achieve a specific purpose. A custom user interface adds functionality to a modelling tool by providing enhanced capabilities. An example of a custom UI is to edit or configure model information that may not be possible using the modelling tool's default functionalities.
	
\end{itemize}

\subsubsection{Usability Heuristics}
Usability heuristics provide guidelines to language engineers in making decisions during graphical DSML development that are ultimately beneficial to users in achieving modelling with a greater sense of effectiveness and satisfaction~\cite{poltronieri2017usability, poltronieri2018usa}.
Usability heuristics, should therefore, complement the earlier proposed design decisions.
Since most graphical modelling tools have an underlying Application Programming Interface (API) to support in the development of DSMLs, we consider the following usability taxonomy attributes, proposed and compared to UX studies in APIs~\cite{mosqueira2020usability}, important in graphical modelling: 
\begin{itemize}
	\itemsep0em
	\item \textbf{Knowability (Clarity) (U-1)}: constructs of the modelling language should be self-explanatory.
	
	\textit{Rationale}: Readability is often a challenge users face during modelling of complex scenarios. A clarity in terms of names, types, structure, logical grouping of common model elements, easy to comprehend modelling constructs, which elements to use where and when, and the ability to design model elements that serve a particular function, is critical to improving the overall modelling experience.
	
	\item \textbf{Knowability (Helpfulness) (U-2)}: the graphical DSML should provide helpful annotations and documentation to users and also identify deprecated elements.
	
	\textit{Rationale}: Documentation is a key aspect often missing when building DSMLs. Users are left to comprehend the meaning and usage of modelling language constructs by themselves, which is tedious and time consuming. Providing sufficient documentation with good usage examples reduces the effort needed to model complex elements and consequently limits introducing errors during modelling.
	
	\item \textbf{Operability (U-3)}: the graphical DSML should provide the necessary domain-specific functionalities in addition to being extensible for further language compositions.
	
	\textit{Rationale}: DSML users should be presented with constructs that are relevant to their domain and performs the specified functionality. Model elements should be precise, universally recognized, and flexible to allow language compositions and extensibility to support effective model transformations.
	
	\item \textbf{Robustness (U-4)}: the graphical DSML should be well-formed and be free from bugs and vulnerabilities that could potentially expose flaws in the system.
	
	\textit{Rationale}: Every DSML should be thoroughly tested to ensure that runtime errors are not encountered during modelling with a graphical modelling tool. The DSML constructs should be error free and checked for potential vulnerability leaks.
	
	\item \textbf{Safety (U-5)}: the graphical DSML should not compromise the confidentiality or the assets of a user.
	
	\textit{Rationale}: Data belonging to the user should not be exposed to unauthorized entities at any cost. License of use, legality, and the personal information of users and their data should be protected to ensure the safety of users' assets.
	
\end{itemize}

\subsection{Scope of UXD}
The overall topic of UXD is very wide and can potentially cover a lot of scenarios.
The design decisions we discuss are non-exhaustive, yet important in the development of any graphical DSML and leads to a good UX.
Further, the combination of quality management standards such as ISO standards along with the ergonomics of human-system interaction provide best practices to improve graphical DSMLs for a seamless UX.
Language engineers should not only focus on the aesthetics of modelling language constructs but also account for design decisions that improve the general usability of graphical modelling languages.
It is also common for language engineers to focus more on the functional aspects of the modelling language.
Constraints in project duration, resources, and budget introduces trade-offs in design decisions that language engineers must consider.
To this end, we propose the following three questions that language engineers must answer during language development:
\begin{itemize}
	\itemsep0em
	\item \textbf{Q1}: Does a specific design decision fulfil a user's needs or a modelling goal?
	\item \textbf{Q2}: Is the design decision a cause for any potential conflict, either between the constructs of the modelling language or with the existing functionalities of the modelling tool?
	\item \textbf{Q3}: Is the design decision specific, non-subjective, and has relevance to the domain in consideration?
\end{itemize}
Answers to the above questions help language engineers define the scope of UXD. 
In principle, those design decisions that are not aligned with the modelling goals should not be considered.
These must be examined with all the stakeholders in the project to avoid compromising the overall quality and UX of a graphical DSML.


\section{Case Study}
\label{sec:impl}

\begin{figure*}[h!tb]
	\centering
	\includegraphics[width=\textwidth]{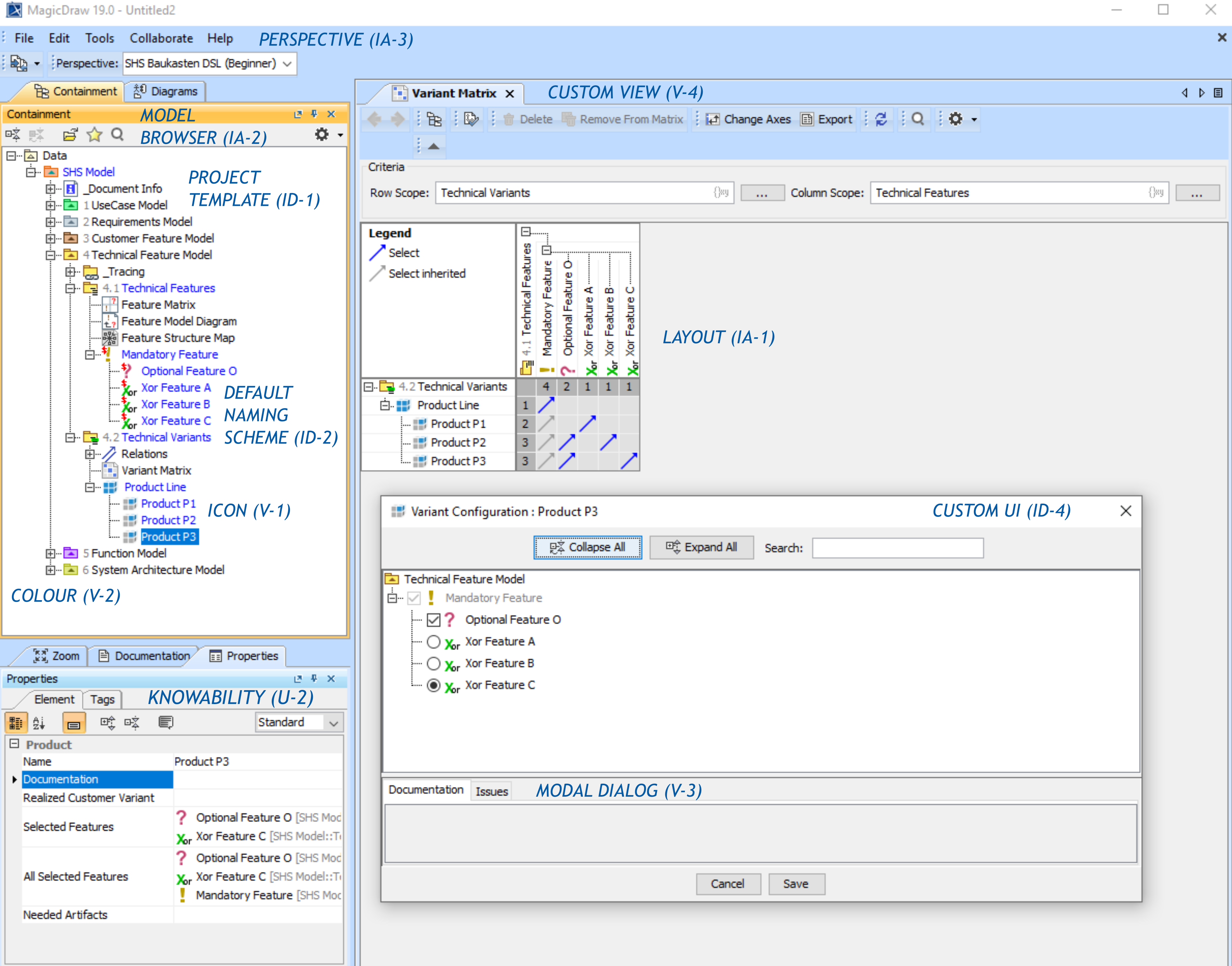}
	\caption{Annotations of design decisions in MagicDraw for a feature model consisting of (1) product lines and products, (2) mandatory and optional features, and (3) a UI for variant configuration.
	The figure shows enhanced aesthetics of models, the structuring and organization of model elements, the interactive aspects focussing on the cognitive dimensions, and the various usability aspects that complement the design decisions.}
	\label{fig:mdExample}
\end{figure*}

To demonstrate the significance and implications of the presented UX guidelines, in this section, we show their effects using a DSML in MagicDraw.
In \autoref{fig:variantConfig}, we presented an example of the configuration of the language elements, product line, and product for a feature model~\cite{BEK+19} DSML that has been developed for Siemens Healthineers.
Further, we reuse the feature model DSL building block, in other DSMLs such as the one we developed for Siemens Digital Industry (\autoref{sec:modInMD}).
The customization capabilities in MagicDraw allow design decisions such as icon (V-1), colour (V-2), default name (ID-2), and creation views (IA-4) to be incorporated directly into the stereotype definitions and their corresponding customizations.
Modal dialogs (V-3) are created using the MagicDraw Open Java API capabilities and custom views (V-4) are created using the customization capabilities offered by MagicDraw.
The information architecture and interaction designs are also configurable using the MagicDraw configuration settings. 
Java based plugin mechanism allow various plugins creation, such as our dynamic view plugin (V-5).
Plugins that enable model transformations (ID-3) and custom UI (ID-4) are also integrated directly into the feature model DSML.

\autoref{fig:mdExample} shows an illustration of a feature model created by a user in MagicDraw and is based on the design decisions defined by the language engineers.
The model browser (IA-2) is a functionality of MagicDraw that hierarchically organizes and structures the model constructs, such that a user can find and navigate quickly through model elements.
A predefined feature model project template (ID-1) is created automatically inside the model browser upon the creation of a new feature model project.
This template structures and organizes the various models configured internally as DSL building blocks (\autoref{sec:modInMD}).
This ensures that all the relevant models and model elements are exclusive to their respective building blocks and eventually help users in quicker navigation and ease of finding models, thus addressing Q1.
The predefined project template includes a default naming scheme (ID-2), with automatic naming and numbering conventions, such as \quotes{1 UseCase Model} and \quotes{2 Requirements Model}, so that users do not forget to add relevant identifiers and are also guided to create models in a sequential manner (U-1).
Each of the model elements is also configured with icons (V-1), and colouring (V-2) of features in the feature model.
While MagicDraw, by default, provides the ability to set icons and colours, it is left to the language engineers to design the appropriate icons and colouring to their model elements that effectively represents the domain in consideration.
In our example, we designed the product line icon to show four blue cubes whereas a product icon, say P1, is configured to show only one blue cube and three other white cubes, thus providing the distinction between a product line and a product.
Similarly, design differences between the icons of different feature types, such as mandatory (with an exclamation mark) and an optional feature (with a question mark), provides the appropriate distinction between the different model elements and helps improve the aesthetics for users during modelling.
The designs for the icons we built for our feature model are similar to the icons used by fully featured product line engineering tools like pure::variant~\cite{beuche2008modeling}.
We also note that novice users may not necessarily be aware of standard representations, and therefore proper documentation (U-2) for such constructs are also needed.

MagicDraw offers perspectives (IA-3), for enabling or disabling certain MagicDraw functionalities that maybe beneficial to advanced users.
In addition, language engineers can use this functionality to restrict certain modelling constructs that is only visible to advanced users.
In our example, the perspective (IA-3) is configured to a beginner (a novice user), showing only basic MagicDraw toolbar options that is a good starting point for novice users.
A perspective for an advanced user allows additional functionalities such as collaboration on cloud servers for project migration, advanced context condition checking, and merge of different projects.
Perspectives do not cause any potential conflicts between the feature model DSML constructs or the existing functionalities of MagicDraw, thus addressing Q2.
The custom variant matrix (V-4) shows various product lines or products and their respective connections to the various features.
This is also supported by specifying a layout in the matrix definition (IA-1) that combines the variants and the features.
In our example, we show the product lines and products in the rows of the matrix, whereas the features are displayed on the columns for the matrix.
The legend information, enabled using our dynamic view plugin (V-5), on the matrix shows the kinds of relationships available for the product lines, products, or features.
This would otherwise be not possible as matrices and tables, in contrast to class diagrams, composite structure diagram, and so on, do not generally allow legends information by default in MagicDraw.
Additionally, the matrix is interactive, meaning users can directly right click on an empty cell and choose a relationship, directed from the variants to the features, thereby addressing Q3.
Using the MagicDraw Open Java API a dedicated variant configuration user interface (ID-4) is shown to users, allowing for a convenient configuration of product lines or products.
Based on the selection of users, we can automatically create or remove relationships between the model elements.
Additional options, such as collapsing and expanding of the hierarchical structure of features, and also the inclusion of a search bar to effectively search for features in a complex or long list of items, enhances the configuration of each variant.
A modal dialog (V-3), integrated with the whole variant configuration UI (ID-4), presents additional documentation and issues that could possibly occur during a variant configuration.
In general, the usability heuristics are applied throughout the DSML.
They ensure that modelling language constructs remain self-explanatory (U-1) and are supported with helpful documentation and annotations (U-2).
Language engineers must also validate and verify the DSML and any external source code, such as GPL code to create user interfaces or dynamic view plugins, to ensure robustness (U-4) and safety (U-5) of the modelling language and the modelling tool.

\begin{figure*}
	\centering
	\includegraphics[width=\textwidth]{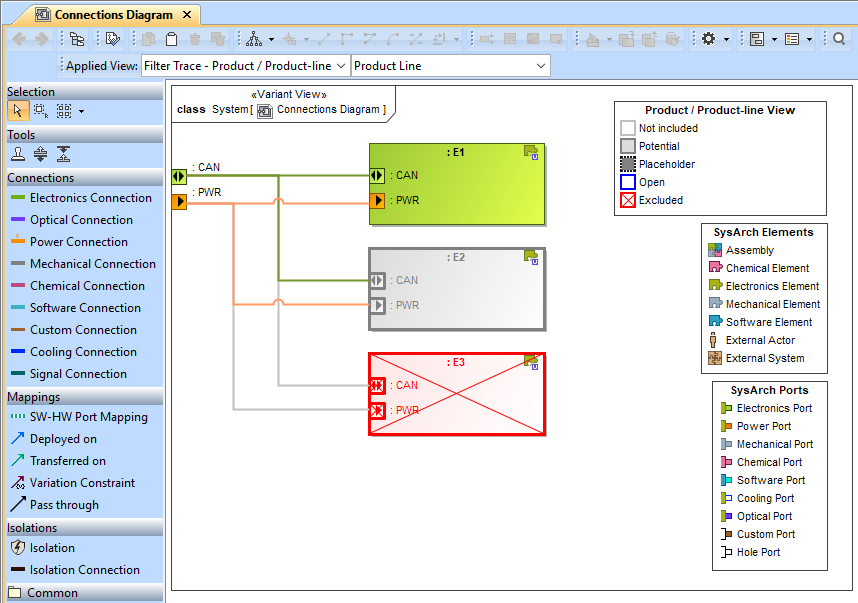}
	\caption{An example of a custom view (V-4) integrated with a dynamic view plugin (V-5), that filters diagram elements based on a configured product line or product. The element E2 is marked as potential, which means that the element might or might not be included in the products, but on the currently displayed product line it is still undecided. On the other hand, the element E3 is marked as excluded, hence the respective connections are greyed out and the element is marked with a red cross.}
	\label{fig:mdExample2}
\end{figure*}

\autoref{fig:mdExample2} shows an illustration of how a custom view (V-4) is integrated with the Java API based dynamic view plugin (V-5) in a feature model to enhance the UX in a graphical environment.
While only the graphical modelling canvas is provided by MagicDraw, the configuration and structuring (IA-1) of model elements inside the canvas is configured by the language engineers.
During design time, users can reposition and restructure these model elements based on their needs, fostering flexibility in their modelling.
The applied view can be dynamically changed to select either a product line or a product in a feature model tree.
The selected configurations for this particular variant, a product line in this case, is dynamically updated and made visible to users in this connections diagram (V-4).
In this example, the electronic element E1 is configured to be included (IA-1) using a custom user interface (ID-4).
Therefore the electronics and power connections to E1 is also made visible to this product line on this custom view.
The element E2 is marked as potential, meaning it is currently not included in the currently displayed product line, but could be included for a subsequent product in the product line.
In this scenario, the respective connections are visible, although the element itself is greyed out for the product line.
Finally, element E3 is marked for exclusion for this product line, meaning the features are excluded for subsequent products.
Such a scenario can exist when building a low cost variant of a product line requiring less features.
For E3, the connections are greyed out and the element is marked with a red cross (V-1, V-2).

\begin{figure}
	\centering
	\includegraphics[width=0.45\textwidth]{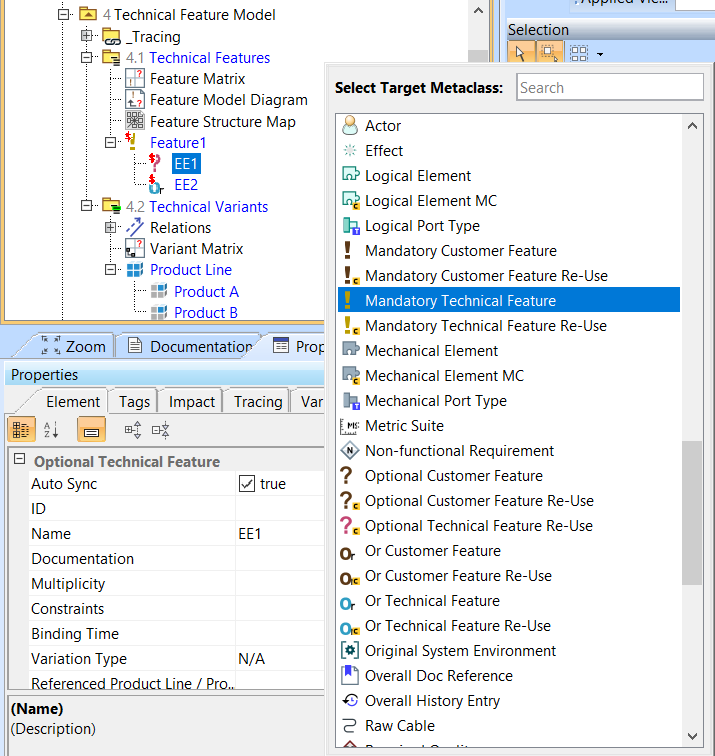}
	\caption{An example of a model transformation (ID-3) in the form of refactoring an optional feature to a mandatory feature. While the modal dialog for selecting the target metaclass lists all possible element conversion types, the risk of losing incompatible properties is informed to users.}
	\label{fig:modTxf}
	\vspace{-5mm}
\end{figure}

Model transformations (ID-3) in the form of refactoring of models, is achieved by changing the stereotype configuration of a model element.
A modal dialog (V-3) is created by language engineers using the MagicDraw customization capabilities to support such a kind of model transformation.
\autoref{fig:modTxf} shows an example of a model transformation (ID-3) for refactoring features in a feature model.
An already defined optional feature in a feature model can be refactored to a mandatory feature.
While such a transformation is allowed during design time to enhance the modelling experience and to introduce extensibility of the feature model language, it also includes the risk of losing certain incompatible properties, which is subsequently informed to users using a modal dialog (V-3).

\begin{figure}
	\centering
	\includegraphics[width=0.45\textwidth]{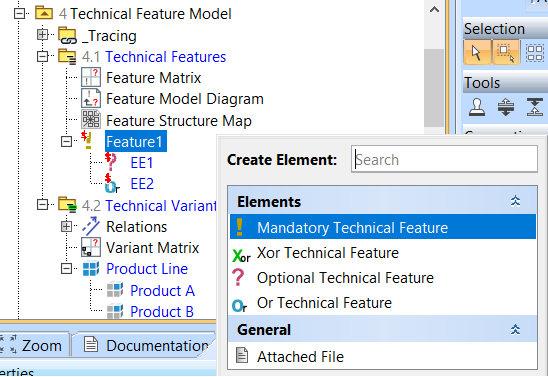}
	\caption{An example illustrating the creation view (IA-4) where the creation of a model element inside a Technical Features package allows only a mandatory, or, Xor, or an optional feature to be created.}
	\label{fig:creationView}
	\vspace{-10mm}
\end{figure}

\autoref{fig:creationView} illustrates an example of a creation view (IA-4), created by language engineers, that lists the creation of only feature elements: mandatory, optional, or, and Xor.
This additional window showing the logical grouping of language elements ensures that users can only create the above mentioned four features under the \quotes{4.1 Technical Features} package.
This avoids inconsistencies in model element creation, such as the creation of features under the \quotes{4.2 Technical Variants} package, which introduces disorder in logically structuring and grouping elements together.
Further, a default naming scheme (ID-2) not only assigns relevant names and labels to features and variants, but also guides a user into first creating the features, within \quotes{4.1 Technical Features} package.
Next, users create the variants, within \quotes{4.2 Technical Variants} package, needed for a product line or a product configuration.

\section{Discussion}
\label{sec:discussion}
In this paper, we present definitions for UX and UXD for graphical DSMLs in MagicDraw.
Further, we discuss a categorized set of design decisions that help improve UX for all kinds of users to achieve their modelling.
Many traditional approaches to developing graphical DSMLs exist, but they often lack good UX aspects that language engineers must consider during DSML development.
The combination of a methodology along with a modelling tool elevates the experience of using graphical DSMLs, which can get complex and hard to use.
In our years of experience in building graphical DSMLs for industrial projects, we have explored MagicDraw as a graphical modelling tool capable of providing the combination of a modelling language, methodology and good UX for both novice and advanced users.
Relevant key stakeholders must be involved from the start of the project.
An iterative feedback loop should be established between users (modellers), domain experts, and language engineers to understand the specifics of the domain in consideration.

Language engineers are certainly not UX or programming experts, therefore, they often need additional trainings.
Consequently, design decisions taken by language engineers are aimed at improving the UX.
Categorization of the design decisions achieves separation of concerns for language engineers possessing different skills.
Visual designs improve the aesthetics of the model elements and the models.
Models should be designed to look and feel similar to their real world abstractions for easy understandability.
Icons, colours, modal dialogs, and custom views can be configured to model real world abstracts.
Information architecture design decisions help organize and structure the constructs of the modelling language so that they are easy to find and use.
Interaction designs help users interact effectively with the modelling language while focussing on the cognitive dimension.
These designs alleviate problems of building models from scratch with little guidance.
Finally, usability heuristics provide the necessary guidelines to language engineers in developing graphical DSMLs that improve the effectiveness and satisfaction of users.
Therefore, we consider the taxonomy attributes of knowability, operability, robustness, and safety most important.

The customization capabilities of MagicDraw allow design decisions to be easily integrated into the DSML.
Each language element can be configured as per the project requirements and language components consisting of such language elements, or stereotypes, can be easily created and bundled together with MagicDraw.
Functionalities that are not provided by default can be achieved through Java based plugin mechanisms in MagicDraw.
Finally, all the components are composed together into the final DSML that is used by modellers to realize their modelling.
Leveraging the customization capabilities of MagicDraw allow language engineers to capture a greater variety of design decisions that ultimately are beneficial for all kinds of users.
Further, building blocks from heterogeneous domains are composed together that ultimately fosters re-usability of language components and design decisions.


Generally, our design guidelines for improving the UX are tailored to industrial DSMLs. 
However, they naturally apply to all graphical DSMLs (e.g., research). 
We specifically focus on industrial languages since they have a greater necessity for good UXD decisions. 
While DSMLs in research often represent proofs of concept that do not need to support all categories of visual design, information architecture, interaction design, or usability heuristics, this is essential for industrial applications as they are intended for practical use. 
Therefore, although the guidelines have a more general claim, they present particular challenges in the industrial context.
Specific guidelines, such as defining a custom UI for generating abstract syntax trees from a textual grammar, could also be considered for non-graphical DSMLs but needs more research.
Additionally, while the presented guidelines are primarily based on our experiences with the modelling tool MagicDraw, they generally apply to comparable frameworks as well, as the principle of graphical modelling remains unchanged.

Naturally, no single solution exists for improving the UX in graphical DSMLs.
This is due to the fact that UX is a vastly subjective topic covering a wide range of domains and is a research area that transcends traditional usability heuristics.
Often, language engineers are not UX experts and struggle to effectively convey the semantics of a modelling language.
While some definitions of UX can be reused in the context of graphical DSMLs, it is often overly generic or tied to very specific technological spaces.
We believe our definitions of UX and UXD along with the categorization of UXD aspects in MagicDraw will foster the discussion in improving UX for graphical DSMLs across various modelling tools.
Language engineers should work cohesively with UX experts in integrating design decisions during industrial DSML development for improving the UX.
Consideration should also be made for language engineers to be trained with relevant domain and UX knowledge.
The aspects of UX described in our paper are not exhaustive, but in our experience of developing industrial DSMLs, ameliorates the challenges faced by a multitude of users in graphical modelling.
While these design aspects may seem specific to MagicDraw as tooling functionalities, we observe that other graphical modelling tools also provide ample functionalities and customization capabilities for the effective application of these design decisions.
We intend on exploring the application of the design aspects considered in this study to other modelling tools as well.
In this way, both the risk to a vendor-locked scenario and the generalizability of the design aspects can be overcome.
This would also help us shape, refine, and improve the design guidelines to be able to perform an elaborate guideline review.
Further, the scope of UX and UXD must be defined by language engineers and trade-offs in design decisions must be carefully analysed according to project requirements, resources, and costs.
One key takeaway in developing industrial DSMLs across various domains over the years is that these guidelines improve the overall DSML experience of users and domain experts in modelling effectively.
To this end, we consider the proposed UX and UXD guidelines in this paper a good reference point for future discussions towards defining UX aspects for industrial DSMLs that is independent of graphical modelling tools.

\vspace{-4pt}
\section{Related Work}
\label{sec:relWork}

Many studies exist on general design guidelines for DSLs and DSMLs~\cite{KKP+09,365788,frank2013domain}.
While these guidelines apply generically to DSLs, our guidelines are tailored specific to graphical modelling tools and focusses on industrial DSMLs designed for practical use.
Studies on usability driven development of DSLs~\cite{barivsic2018usability,borum2021designing,poltronieri2018usa} focus on usability evaluation during the DSL development process to address usability problems, but miss discussions on a variety of complex industrial domains.
In contrast, our guidelines have been applied over the years on a wide variety of industrial domains (\autoref{sec:modInMD}).
Often, experimental usability evaluations~\cite{karna2009evaluating} do not consider all stages in the DSML development lifecycle.
Our guidelines propose that the design decisions are taken early on in the projects with all the stakeholders involved.
While crowdsourcing and collaborative techniques for shaping graphical notations of DSLs have emerged~\cite{brambilla2017better,izquierdo2016collaboro}, it allows for more subjectivity as it permits a wider spectrum of users to collaborate, validate, and promote DSL acceptance.
For the description of interaction designs in our guidelines, the cognitive dimensions of notations framework (CDF) \cite{green1989cognitive,blackwell2001cognitive} has been considered.
This is beneficial in covering cognitive dimensions aspects of \textit{hidden dependencies}, \textit{diffuseness}, and \textit{viscosity}.
While human-centred design approaches for usability evaluation~\cite{poltronieri2017usability} have been reviewed, we consider the principles of human-centred design defined in ISO 9241-210~\cite{ISO9241210} to be generally applicable.
In our work, we consider only certain DSL usability heuristics proposed by~\cite{mosqueira2020usability} important in the context of industrial DSMLs.
Further, challenges and future direction for UX in model-driven engineering approaches is discussed in~\cite{abrahao2017user}, whereas our work focusses on improving the general usability and UX in industrial graphical DSMLs.

\vspace{-4pt}
\section{Conclusions}
\label{sec:conclusions}

As systems become complex and heterogeneous in nature, challenges in developing industrial DSMLs that ameliorate UX have emerged.
Modelling tools such as MagicDraw aim to improve UX by introducing a variety of customization capabilities during graphical DSML development.
While such tools provide sufficient functionalities for developing a modelling language, there still exists the challenge of providing UX and UXD guidelines for language engineers.
To address this challenge, we leverage the standards of human-centred design and UX in general, and propose aspects of UX and UXD for graphical modelling languages in MagicDraw, which we hope is generalizable to other graphical modelling tools.
We categorize design decisions by utilising the wide range of customization capabilities in MagicDraw: visual designs to improve the aesthetics of model and model elements, information architecture to organize and structure the constructs of the modelling language, interaction designs to help users interact seamlessly with the modelling language and providing usability heuristics language engineers should consider to help users model with effectiveness and satisfaction.
These aspects of UX and UXD improve the overall experience of novice and advanced users in using graphical DSMLs.
Further, these guidelines serve as a reference point for future discussions towards defining UX aspects for graphical DSMLs.
Naturally, UX is a subjective topic and our proposed list of design decisions are non-exhaustive.
We consider discussions on improving UX for graphical modelling for every kind of user important.
We believe guidance for language engineers in developing industrial DSMLs that combines various UX aspects ultimately improves the usability and UX
in graphical modelling that is independent of graphical modelling tools.


\Urlmuskip=0mu plus 1mu\relax

\bibliographystyle{ACM-Reference-Format}
\bibliography{bib,se}

\end{document}